\documentclass[12pt,oneside]{article}
\usepackage{amsfonts,amssymb,graphicx}
\def\qed{\par\noindent\rightline{$\square$}}
\def\({\left (}
\def\){\right)}
\def\[{\left [}
\def\]{\right]}
\def\la{\langle}
\def\ra{\rangle}

\def\be{\begin{equation}}
\def\ee{\end{equation}}
\def\bea{\begin{eqnarray}}
\def\eea{\end{eqnarray}}
\def\<{\langle}
\def\>{\rangle}
\def\~{\tilde}
\def\s{\sigma}

\def\a{\alpha}
\def\b{\beta}
\def\g{\gamma}
\def\e{\epsilon}

\newcommand{\E}{\Bbb E}
\newcommand{\R}{\Bbb R}

\newcommand{\C}{\Bbb C}

\newcommand{\Z}{\Bbb Z}

\newtheorem{theorem}{Theorem}
\newtheorem{lemma}{Lemma}
\newtheorem{definition}{Definition}
\newtheorem{corollary}{Corollary}
\newenvironment{proof}{{\bf Proof:\ }}{\hfill$\square$\vskip.5cm}
\newcommand{\nl}{\newline}
\newcommand{\non}{\nonumber}
\def\Av{{\rm Av}}
\def\Tr{{\rm Tr}\,}
\def\half{\frac{1}{2}}
\begin{document}
%
\noindent RMP-818 09-04-04 \vskip 1.5truecm
\begin{center}
{\sc\Large
the thermodynamic limit for finite dimensional
classical and quantum disordered systems
}
\vskip 1cm
{\bf
Pierluigi Contucci\footnote{contucci@dm.unibo.it},
Cristian Giardin\`a\footnote{giardina@dm.unibo.it}
}\\
{\small Dipartimento di Matematica} \\
{\small Universit\`a di Bologna,
40127 Bologna, Italy}
\vskip .24truecm
{\bf
and
\\
Joseph Pul\'e\footnote{Joe.Pule@ucd.ie}}\\
{\small Department of Mathematical Physics}\\
{\small University College Dublin, Belfield, Dublin 4, Ireland}
\end{center}
\vskip 1truecm
\begin{abstract}\noindent
We provide a very simple proof for the existence of the thermodynamic limit for the quenched specific pressure
for classical and quantum disordered systems on a $d$-dimensional lattice, including spin glasses. We
develop a method which relies simply on Jensen's inequality and
which works for any disorder distribution with the only condition (stability)
that the quenched specific pressure is bounded.
\end{abstract}
\newpage
\section{Introduction, definitions and results}
In this paper we study the problem of the existence of the thermodynamic
limit for a wide class of disordered models defined on finite dimensional
lattices. We consider both the classical and quantum case with random
two-body or multi-body interaction. The classical case has been studied in various places
(see for example \cite{V, PF, KS,  VV} and \cite{Z}). In \cite{V} and \cite{VV} the quantum case with pair interactions
has also been considered. Here we deal only with the quenched pressure and
using only thermodynamic convexity and a mild stability condition
we give a very simple proof of the existence and monotonicity of the quenched specific pressure.
A result in the same spirit for classical spin glasses has been obtained in \cite{CG} by using an interpolation technique
introduced in \cite{GT, CDGG}. The present work extends the results of \cite{CG} not only
to the quantum case, but also to the classical case with a non zero mean of the interaction and to 
the continuum spin space.
\\
We shall treat the classical and quantum cases in parallel.
In the classical case to each point of the lattice $i\in\Z^d$ we associate a copy of the {\it spin space} ${\cal S}$,
which is equipped with an a priori probability measure $\mu$. We shall denote this by
${\cal S}_i$. In the quantum analogue we associate to each $i\in\Z^d$ a copy of  a finite dimensional Hilbert space ${\cal H}$, denoted by ${\cal H}_i$ and
a set of self-adjoint operators, {\it spin operators}, on ${\cal H}_i$.
\\
Following \cite{Ru}, (see also \cite{S}), we define the interaction in the following way. In the classical case for each finite subset of $\Z^d$, $X$, we let
${\cal S}_X := \times_{i \in X}{\cal S}_i$ and $\{\Phi^{(j)}_X\, |\, j\in n_X\}$ is a finite set of bounded functions from ${\cal S}_X$ to $\R$ which are measurable
with respect to the product measure $\mu^{|X|}$ on ${\cal S}_X$.
In the quantum case each $\Phi^{(j)}_X$
is a self-adjoint element of the algebra generated by the set of operators, the {\it spin operators} on ${\cal H}_X := \otimes_{i \in X}{\cal H}_i$.
Without loss of generality we set $\Phi_{\emptyset}=0$.
In both cases we take the interaction to be translation invariant in the sense that if $\tau_a$ is translation by $a\in \Z^d$, then
\be
n_{\tau_aX} =n_X\ \ \ {\rm and}\ \ \ \Phi^{(j)}_{\tau_aX} =\tau_a \Phi^{(j)}_X\ \ {\rm for}\ j\in n_X.
\ee
We now define the random coefficients. For each $X$ let $\{J^{(j)}_X\, |\, j\in n_X\}$ be a set of random variables.
We assume that the $J^{(j)}_X$\,'\,s are independent random variables and that $J^{(j)}_{\tau_aX}$ and
$J^{(j)}_X$ have the same distribution for all $a\in \Z^d$.
We shall denote the average over the $J$'s by $\Av [\cdot]$.
\vskip 0.5 cm \noindent
Let $\Lambda\subset \Z^d$ be a finite set of a regular lattice in
$d$ dimensions and denote by $|\Lambda|=N$ its cardinality. We
define the {\it random potential} as
\be
U_\Lambda(J,\Phi) :=
\sum_{X\subset\Lambda}\sum_{j\in n_X}J^{(j)}_X\Phi^{(j)}_X .
\label{sgp}
\ee
We stress here that the distributions of the
$J^{(j)}_X$'s are independent of the volume $\Lambda$. This
characterizes the short range case, such as the Edwards-Anderson
model. In mean field (long range) models, such as the
Sherrington-Kirkpatrick model in which the Hamiltonian sums over
all the couples ($N^2$ terms), the variance of $J^{(j)}_X$ has to
decrease like $N^{-1}$ in order to have a well defined thermodynamic
behaviour and in particular a finite energy density.\\
The complete definition of the model we are considering requires
that we specify also the interaction on the frontier $\partial \Lambda$,
i.e. boundary conditions. However, standard surface over volume
arguments imply that if the quenched specific pressure for one boundary condition converges,
then it also converges for all other boundary conditions. Therefore to prove the convergence of the
quenched specific pressure it is sufficient to consider the free boundary condition.
Thus in the sequel we shall assume the free boundary condition and prove that in this case the quenched
pressure is monotonically increasing in the volume. \\\\
We would like to emphasize the fact that
in the classical case, our results are not restricted to the situation when the space ${\cal S}$
consists of a finite number of points. Here we also want to cover the case of continuos spins and therefore we
shall keep the classical and quantum cases separate. Of course both cases can be covered simultaneously in a
C$^*$ algebra setting but for the sake of simplicity we shall not take this route.\\
\par
\noindent
{\bf Examples}:
\begin{enumerate}
\item {\it Classical Edwards-Anderson model}
\newline
${\cal S} = \{-1,1\}$,
$\mu(\s_i) = \frac{1}{2}\delta(\s_i + 1) +
\frac{1}{2}\delta(\s_i - 1)$. The interaction
is only between nearest neighbours:
$\Phi_{i,j}(\s_i,\s_j) = \s_i\s_j$ for
$|i-j|=1$, $\Phi_X = 0$ otherwise.  To ensure that the specific pressure
is bounded it is enough that
\be
\Av\[|J_{ij}|\]<\infty.
\ee
More generally one may consider
a long range interaction with $\Phi_{i,j}(\s_i,\s_j) = \s_i\s_j/R(|i-j|)$
with a sufficient condition for boundedness, for example
\be
\Av\[J_{0i}\]=0\ \ \ {\rm and} \ \ \ \sum_i\frac {\Av\[|J_{0i}|^2\]}{\(R(|i|)\)^2}<\infty,
\ee
or a many-body interaction with a suitable decay law. One can also add a (random) external field.
\nl
We refer the reader to \cite{CG} for more classical examples.
\item {\it Quantum Edward-Anderson model}
\newline
${\cal H} = \C^2$.
The spin operators are the set of the Pauli matrices:
$\s_i = (\s_i^x,\s_i^y,\s_i^z)$
\be
\sigma^x =
\left( \begin{array}{cc} 0 & 1 \\ 1 & 0 \end {array} \right)
\qquad
\sigma^y =
\left( \begin{array}{cc} 0 & -i \\ i & \phantom{-}0 \end {array} \right)
\qquad
\sigma^z =
\left( \begin{array}{cc} 1&\phantom{-}0\\0&-1 \end {array} \right)
\qquad
\ee
which commutation and anticommutation relations
\be
[\,\s_i^{\a},\s_i^{\b}\,] = 2 i \e_{\a \b \g} \,\s_i^{\g}
\ee
\be
\{\s_i^{\a},\s_i^{\b}\} = 2 \delta_{\a \b}
\ee
The interaction is again only between nearest
neighbours: $\Phi_{i,j}(\s_i,\s_j) = \s_i\cdot \s_j =
\s_i^x\,\s_j^x + \s_i^y\,\s_j^y + \s_i^z\,\s_j^z$
for $|i-j|=1$, $\Phi_X = 0$ otherwise. A transverse
field $\Phi_i(\s_i) = \s_i^z$ can also be added. One can have an asymmetric version with local interaction
\be
J^x_{i,j}\Phi^x_{i,j}(\s_i,\s_j)+J^y_{i,j}\Phi^y_{i,j}(\s_i,\s_j)+J^z_{i,j}\Phi^z_{i,j}(\s_i,\s_j)
\ee
where
$\Phi^x_{i,j}(\s_i,\s_j) = \s_i^x\,\s_j^x$, $\Phi^y_{i,j}(\s_i,\s_j) = \s_i^y\,\s_j^y$ and
$ \Phi^z_{i,j}(\s_i,\s_j) = \s_i^z\,\s_j^z$.
As in Example 1 one may consider
a short range interaction with a suitable decay law.
\end{enumerate}
{\bf Notation:}
We shall use the notation $\Tr$ to denote both
the classical expectation over ${\cal S}^N$
with the measure $\mu(d\s)=\prod_{i=1}^N\mu(d\s_i)$
and the usual trace in quantum mechanics on the
Hilbert space $\otimes_{i=1}^N {\cal H}$.
\begin{definition}
We define in the usual way:
\begin{enumerate}
\item The random partition function, $Z_\Lambda(J)$, by
\be
Z_\Lambda(J) \, := \, \Tr e^{U_\Lambda(J,\Phi)}
\; ,
\ee
\item The quenched pressure, $P_\Lambda$, by
\be
P_\Lambda := \, \Av[\, \ln Z_\Lambda(J)\, ] \; ,
\label{fe}
\ee
\item The quenched specific pressure, $p_\Lambda$, by
\be
p_\Lambda :=\frac{P_\Lambda}{N} \; .
\label{fe}
\ee
\end{enumerate}
\end{definition}
We are now ready to state our main theorem:
\begin{theorem}
\label{teo}
If all the $J^{(j)}_X$\,'s with $|X|>1$ have zero mean
then the quenched pressure is superadditive:
\be
P_\Lambda \, \ge \,  \sum_{s=1}^{n}P_{\Lambda_s} \; .
\label{super}
\ee
\end{theorem}
Let $\|\Phi^{(j)}_X\|$ denote the supremum norm in the
classical case and the operator norm in quantum case.
For the case when the $J^{(j)}_X$\,'s do not have zero mean we have the following corollary:
\begin{corollary}
\label{cor}
Let
\be
{\bar P}_\Lambda=P_\Lambda+\sum_{X\subset\Lambda,\, |X|>1}\sum_{j\in n_X}|\Av [J^{(j)}_X]|\,\|\Phi^{(j)}_X\|.
\ee
Then ${\bar P}_\Lambda$ is superadditive.
\end{corollary}
Theorem \ref{teo} combined with the boundedness of the specific
pressure is sufficient to ensure the convergence of the specific
pressure in the thermodynamic limit (see for example \cite{Ru}
Chapter IV) in the case when all the $J^{(j)}_X$\,'s with $|X|>1$ have zero mean.
In the case when the $J^{(j)}_X$\,'s do not have zero mean
we have to add to Corollary \ref{cor} the condition
\be
C:=\sum_{X\ni 0,\, |X|>1}\sum_{j\in n_X}\frac{|a^{(j)}_X|\|\Phi^{(j)}_X\|}{|X|}<\infty.
\label{C}
\ee
This implies that
\be
\lim_{\Lambda\to\infty}\frac{1}{N}\sum_{X\subset\Lambda,\, |X|>1}\sum_{j\in n_X}|a^{(j)}_X|\|\Phi^{(j)}_X\|=C
\ee
and therefore the convergence of the specific pressure
\par \noindent
To prove the boundedness of the specific pressure we
need the following stability condition (cf \cite{Z}). Let
\be
\|U\|_1:=\sum_{X\ni 0}\sum_{j\in
n_X}\frac{\Av\[|J^{(j)}_X|\]\|\Phi^{(j)}_X\|}{|X|}
\ee and
\be
\|U\|_2:=\(\sum_{X\ni 0}\sum_{j\in
n_X}\frac{\Av\[|J^{(j)}_X|^2\]\|\Phi^{(j)}_X\|^2}{|X|}\)^\half.
\ee
\begin{definition}
We shall say that the random potential $U(J,\Phi)$ is stable if it is of the form
\be
U_\Lambda(J,\Phi)={\tilde U}_\Lambda({\tilde J},{\tilde \Phi})+{\hat U}_\Lambda({\hat J},{\hat \Phi})
\ee
where all the ${\tilde J}^{(j)}_X$\,'s and ${\hat J}^{(j)}_X$\,'s are independent, the ${\hat J}^{(j)}_X$\,'s have zero mean and
$\|{\tilde U}\|_1$ and $\|{\hat U}\|_2$ are finite.
\end{definition}
With this definition we shall prove in the next theorem that the specific pressure is bounded.
Note that the stability condition in Definition 2 implies that $C$ as defined in (\ref{C}) is finite since
$C\leq \|U\|_1$.
\begin{theorem}
For a stable random potential the quenched specific pressure is bounded.
\end{theorem}
In the next section we prove the theorems.
\section{Proof of the Theorems}
We start with the following definition.
\begin{definition}
Consider a {\it partition} of $\Lambda$ into $n$ non empty disjoint sets
$\Lambda_s$:
\be
\Lambda=\bigcup_{s=1}^{n}\Lambda_s \; ,
\ee
\be
\Lambda_s\cap\Lambda_{s'} = \emptyset \; .
\ee
For each partition the potential generated by all interactions among
different subsets is defined as
\be
{\tilde U_\Lambda} := U_\Lambda - \sum_{s=1}^{n}U_{\Lambda_s} \; .
\ee
\end{definition}
>From (\ref{sgp}) it follows that \be {\tilde U_\Lambda} =
\sum_{X\in {\cal C}_\Lambda}\sum_{j\in n_X} J^{(j)}_X\Phi^{(j)}_X \ee where ${\cal
C}_\Lambda$ is the set of all $X\subset\Lambda$ which are not
subsets of any $\Lambda_s$.
\\
The idea here is to eliminate ${\tilde U_\Lambda}$ from the partition function.
We shall use the following three lemmas.

\begin{lemma}
Let $X_1,\ldots X_n$ be independent random variables with zero mean.
Let $F:\R^n\mapsto \R$ be such that for each $i=1,\ldots,n$
$x_i \mapsto F(x_1,\ldots x_n)$ is convex, then
\be
\E\[F(X_1,\ldots X_n)\]\geq F(0,\ldots 0)
\ee
where $\E$ denotes the expectation with respect to $X_1,\ldots X_n$.
\end{lemma}
\begin{proof}
This follows by applying Jensen's Inequality to each $X_i$ successively.
\end{proof}
The following two lemmas are related to the thermodynamic convexity of the pressure.
\begin{lemma}
\label{lemma-classic}
Let $\mu$ be a probability measure on a space $\Omega$, and let $A$ and $B_1,\ldots, B_n$ be measurable real valued functions on $\Omega$.
Then
\be
\E\[\log \int_\Omega
\exp\left\{A(\sigma)+ \sum_{i=1}^n X_i B_i(\s)\right\} \,
\mu(d\sigma)\]
\geq
\log \int_\Omega \exp[A(\sigma)]\mu(d\sigma).
\ee
\end{lemma}
\begin{proof}
We just have to check that if
$$
F(x_1,\ldots x_n)=\log \int_\Omega
\exp\left\{A(\sigma)+ \sum_{i=1}^n x_i B_i(\s)\right\} \,
\mu(d\sigma)
$$
then $x_i \mapsto F(x_1,\ldots x_n)$ is convex. Let
\be
\la C \ra :=\frac{\int_\Omega C(\sigma)
\exp\left\{A(\sigma)+ \sum_{i=1}^n x_i B_i(\s)\right\} \,
\mu(d\sigma)}
{\int_\Omega
\exp\left\{A(\sigma)+ \sum_{i=1}^n x_i B_i(\s)\right\} \,
\mu(d\sigma)}.
\ee
Then, computing the derivatives, we have
\be
\frac{\partial F}{\partial x_i}=\la B_i\ra
\ee
and
\be
\frac{\partial^2 F}{\partial x_i^2}=\left \la B_i^2 \right \ra- \la B_i\ra^2=\left \la \(B_i- \la B_i\ra\)^2 \right \ra\geq 0.
\ee
\end{proof}
The next lemma is the quantum analogue of the previous one.

\begin{lemma}
\label{lemma-quantum}
Let ${\cal H}$ be finite-dimensional Hilbert space,
and let $A$ and $B_1,\ldots, B_n$ be self-adjoint operators on ${\cal H}$.
Then
\be
\E\[\log \Tr \exp(A + \sum_{i=1}^n X_i B_i)\]
\geq
\log \Tr\exp A.
\ee
\end{lemma}
\begin{proof}
Again we just have to check that if
$$
F(x_1,\ldots x_n)=\log \Tr \exp(A + \sum_{i=1}^n x_i B_i),
$$
then $x_i \mapsto F(x_1,\ldots x_n)$ is convex.
The first derivative gives
\be
\frac{\partial F}{\partial x_i}=\la B_i\ra
\ee
where
\be
\la C \ra :=\frac{\Tr Ce^{-H}}{\Tr e^{-H}}.
\ee
with
$$
-H=A + \sum_{i=1}^n x_i B_i
$$
while, for the second derivative, we have
\be
\frac{\partial^2 F}{\partial x_i^2}=\( B_i,  B_i\)- \la B_i\ra^2
\ee
where $\(\cdot,\cdot\)$ denotes the Du Hamel inner product (see for example \cite{S}):
\be
\(C,D\):= \frac{\Tr \int_0^1 ds\ e^{-sH}C^* e^{(1-s)H}D}{\Tr e^{-H}}.
\ee
By using the fact that
$\(C,1\)=\overline{\la C\ra}$ and $\(1,D\)=\la D\ra $
we see that
\be
\frac{\partial^2 F}{\partial x_i^2}=\( B_i- \la B_i\ra,  B_i- \la B_i\ra\)\geq 0.
\ee
\end{proof}

\noindent
{\bf Proof of Theorem \ref{teo}}
\nl
Let us assume first that all the $J^{(j)}_X$\,'s with $|X|>1$ have zero mean.
\bea
P_{\Lambda}  & = & \Av \[\, \ln \Tr \exp U_\Lambda \, \] \nonumber\\
             & = & \Av \[\, \ln \Tr
\exp\(\sum_{s=1}^{n}U_{\Lambda_s} + \sum_{X\in {\cal C}_\Lambda}\sum_{j\in n_X} J^{(j)}_X\Phi^{(j)}_X \)\,\]
\eea
Note that ${\cal C}_\Lambda$ does not contain any $X$ with $|X|=1$. Applying Lemma \ref{lemma-classic} (resp. Lemma \ref{lemma-quantum})
for the classical (resp. quantum) case with
$A  = \sum_{s=1}^{n}U_{\Lambda_s}$ and $B_i = \Phi^{(j)}_X$,
$n=\sum_{X\in {\cal C}_\Lambda}n_X$ we get
\be
P_{\Lambda}   \geq  \Av\[\, \ln \Tr \exp\(\sum_{s=1}^{n}U_{\Lambda_s}\) \]=\sum_{s=1}^{n}\Av\[\, \ln \Tr \exp U_{\Lambda_s} \]
=\sum_{s=1}^{n}P_{\Lambda_s} \; .
\ee
\qed
\par\noindent
{\bf \bf Proof of Corollary \ref{cor}}
\nl
Here we relax the condition that all the $J$'s have zero mean. Let $a^{(j)}_X:=\Av\[J^{(j)}_X\]$ and ${\bar J}^{(j)}_X:=J^{(j)}_X-a^{(j)}_X$
for $|X|>1$ so that ${\bar J}^{(j)}_X$ has zero mean and ${\bar J}^{(j)}_X:=J^{(j)}_X$ if $|X|=1$. Let
\be
U^{(1)}_\Lambda(J,\Phi) :=
\sum_{X\subset\Lambda}\sum_{j\in n_X}{\bar J}^{(j)}_X\Phi^{(j)}_X,
\ee
\be
U^{(2)}_\Lambda(J,\Phi) :=
\sum_{X\subset\Lambda,\, |X|>1}\sum_{j\in n_X}\(a^{(j)}_X\Phi^{(j)}_X+|a^{(j)}_X|\|\Phi^{(j)}_X\|\)
\ee
and
\be
{\bar U}_\Lambda(J,\Phi) :=U^{(1)}_\Lambda(J,\Phi)+U^{(2)}_\Lambda(J,\Phi).
\ee
Then
\be
{\bar  U}_\Lambda(J,\Phi)=U_\Lambda(J,\Phi) +\sum_{X\subset\Lambda,\, |X|>1}\sum_{j\in n_X}|a^{(j)}_X|\|\Phi^{(j)}_X\|.
\ee
Thus ${\bar P}_{\Lambda}$ is the pressure corresponding to ${\bar  U}_\Lambda(J,\Phi)$.
One can then see that ${\bar P}_{\Lambda}$ is super-additive by treating the terms in $ U^{(1)}_\Lambda(J,\Phi)$ as before since each ${\bar J}^{(j)}_X$ has zero mean,
except possibly if $|X|=1$,
and by using the fact that all the terms in $ U^{(2)}_\Lambda(J,\Phi)$ are positive (cf \cite{S}). In the quantum case we need the inequality
\be
\Tr e^{(A+B)}\geq \Tr e^A
\ee
if $B$ is a positive operator.
\qed

\noindent
{\bf Proof of Theorem 2}
\nl
For the proof  in the classical case is given in \cite{Z}. Here we modify that proof to cover the quantum case.
>From the Bogoliubov inequality
\be
\frac{\Tr (A-B)\,e^B}{\Tr e^B}\leq \ln \Tr e^A-\ln \Tr e^B \leq \frac{\Tr (A-B)\,e^A}{\Tr e^A}
\ee
with $A=U_\Lambda(J,\Phi)$ and $B=0$ we get
\bea
\log Z_\Lambda(J)-N\log {\rm dim}{\cal H}
&\leq &\frac{\Tr U_\Lambda(J,\Phi)\,e^{U_\Lambda(J,\Phi)}}{\Tr e^{U_\Lambda(J,\Phi)}}\non\\
&=&\frac{\Tr {\tilde U}_\Lambda({\tilde J},{\tilde \Phi})\,e^{U_\Lambda(J,\Phi)}}{\Tr e^{U_\Lambda(J,\Phi)}}
+\frac{\Tr {\hat U}_\Lambda({\hat J},{\hat \Phi})\,e^{U_\Lambda(J,\Phi)}}{\Tr e^{U_\Lambda(J,\Phi)}}\non\\
&\leq &\|{\tilde U}_\Lambda({\tilde J},{\tilde \Phi})\|
+\frac{\Tr {\hat U}_\Lambda({\hat J},{\hat \Phi})\,e^{U_\Lambda(J,\Phi)}}{\Tr e^{U_\Lambda(J,\Phi)}}.
\eea
Now
\be
\Av\[\|{\tilde U}_\Lambda({\tilde J},{\tilde \Phi})\|\]\leq N\|{\tilde U}({\tilde J},{\tilde \Phi})\|_1 .
\ee
For the other term we use the identity for $A$ and $B$ self-adjoint
\be
\frac{\Tr A \,e^{A+B}}{\Tr e^{A+B}}-\frac{\Tr A\, e^B}{\Tr e^B}=\int_0^1 dt\, (A-\la A \ra_t,\, A-\la A \ra_t)_t
\ee
where $\la \cdot \ra_t$ and $(\cdot,\cdot)_t$ denote the mean and the Du Hamel inner product respectively with respect to $H=-(tA+B)$.
The Du Hamel inner product satisfies
\be
(C,C)\leq \half \la C^*C+CC^*\ra^\half \leq \|C\|^2.
\ee
Therefore
\be
\frac{\Tr A \,e^{A+B}}{\Tr e^{A+B}}-\frac{\Tr A\, e^B}{\Tr e^B}\leq 4\|A\|^2.
\ee
With $A={\hat J}^{j}_X{\hat \Phi}^{j}_X$ and $B=U_\Lambda(J,\Phi)-{\hat J}^{j}_X{\hat \Phi}^{j}_X$ we get
\bea
\frac{\Tr {\hat U}_\Lambda({\hat J},{\hat \Phi})\,e^{U_\Lambda(J,\Phi)}}{\Tr e^{U_\Lambda(J,\Phi)}}
&=&\sum_{X\subset\Lambda}\sum_{j\in {\hat n}_X} \frac{\Tr {\hat J}^{j}_X{\hat \Phi}^{j}_X\,e^{U_\Lambda(J,\Phi)}}{\Tr e^{U_\Lambda(J,\Phi)}}\non\\
& \leq &
\sum_{X\subset\Lambda}\sum_{j\in {\hat n}_X}\Tr {\hat J}^{j}_X{\hat \Phi}^{j}_X\,\frac{e^{U_\Lambda(J,\Phi)-{\hat J}^{j}_X{\hat \Phi}^{j}_X}}{\Tr e^{U_\Lambda(J,\Phi)-{\hat J}^{j}_X{\hat \Phi}^{j}_X}}
\non\\
&&{\hskip 2cm}+4\sum_{X\subset\Lambda}\sum_{j\in {\hat n}_X} | {\hat J}^{j}_X|^2\|{\hat \Phi}^{j}_X\|^2.
\non\\
\eea
Thus since $U_\Lambda(J,\Phi)-{\hat J}^{j}_X{\hat \Phi}^{j}_X$ is independent of ${\hat J}^{j}_X$ and $\Av\[{\hat J}^{j}_X\]=0$,
\be
\Av\[\frac{\Tr {\hat U}_\Lambda({\hat J},{\hat \Phi})\,e^{U_\Lambda(J,\Phi)}}{\Tr e^{U_\Lambda(J,\Phi)}}\]
\leq
4\sum_{X\subset\Lambda}\sum_{j\in {\hat n}_X} \Av\[| {\hat J}^{j}_X|^2\]\|{\hat \Phi}^{j}_X\|^2\leq 4N\|{\hat U}({\hat J},{\hat \Phi})\|^2_2.
\ee
Therefore
\be
P_\Lambda\leq N(\log {\rm dim}{\cal H}+\|{\tilde U}({\tilde J},{\tilde \Phi})\|_1+4\|{\hat U}({\hat J},{\hat \Phi})\|^2_2).
\ee
\qed
%
%
\par
\noindent
{\bf Acknowledgments}
\nl
The authors wish to thank B. Nachtergaele and A. van Enter for very constructive suggestions.
JVP wishes to thank the Department of Mathematics of the University of Bologna, Italy,
for their kind hospitality and University College Dublin for the award of a President's
Fellowship. PC  and CG wish to thank S. Graffi for some very useful discussions.

\end{document}